\documentclass{article}

\usepackage{graphicx} 
\usepackage{subcaption}
\usepackage{color}

\title{An Alternative to Quantum Theory}
\author{
Julian Barbour\thanks{julian.barbour@physics.ox.ac.uk}~~and Maria I. R. Lourenço\thanks{fc56407@alunos.fc.ul.pt}
}
\date{}

\begin{document}

\maketitle

\begin{abstract}
Despite the remarkable empirical success of quantum theory, the interpretation of measurement and the nature of entanglement correlations remain open questions. We outline an alternative framework that seeks to recover the empirical results of quantum theory from a holistic, scale-invariant ontology. Assuming only point particles in Euclidean space, characterised by ratios of masses (and charges) and by the ratios of their mutual separations, the theory describes shapes through a single scale-invariant function; no dynamical evolution is required. In principle, this framework provides a unified description of finite subsystems embedded in a configuration of infinitely many particles and offers a possible route to deriving the observed structures of atoms, molecules and macroscopic bodies without invoking standard quantum mechanics.
\end{abstract}

\section{Introduction}

A striking feature of the universe is the diversity of shapes exhibited by objects of the same kind, such as trees or animals of a given species. The ancient Greek atomists were among the first to seek a systematic explanation for this diversity, proposing that the characteristic shapes of macroscopic objects arise from the arrangement of atoms, which themselves have different shapes and relative sizes.

However, the subsequent development of dynamics shifted the focus of physical theory towards the description of change in time. In particular, the astronomical theories that emerged in antiquity provided an important foundation for the later development of Newtonian dynamics. Yet many of the regularities that physics seeks to explain can also be formulated directly in terms of spatial configurations and the relations between them. This raises the possibility that the explanation of structures observed in the present through their evolution from the past may instead be formulated in terms of shapes and timeless correlations.

This possibility is also suggested by approaches to quantum gravity in which time does not appear as an explicit variable. The Wheeler-DeWitt equation, for example, is formulated without an external time parameter and assigns amplitudes or probabilities to configurations. If configurations are regarded as shapes with an overall scale, however, their absolute size is itself problematic: there is no external ruler against which the size of the universe can be measured. Ratios of lengths within a configuration, by contrast, are independent of the choice of units and provide a natural basis for a relational description of the universe.

In this paper, we investigate the possibility of constructing a theory in which such shapes are the primary objects of study. We take three-dimensional scale invariance in Euclidean space to be the most fundamental symmetry underlying physical theory and ask whether the observed structures of the universe can be explained within this framework. The scale-invariant foundation of the theory makes it intrinsically holistic: rather than describing the evolution of isolated systems in time, it constrains complete configurations and the relations between their constituent parts. Our aim is not to challenge the remarkable empirical successes of established theories, but to ask whether the same empirical facts can be accounted for with fewer assumptions. Physics may have accumulated conceptual layers that are not all necessarily required to account for the observed phenomena. The empirical success of quantum theory, for instance, does not necessarily imply that its mathematical formalism provides a faithful description of nature at its most fundamental level. We therefore seek a minimal ontology capable, in principle, of reproducing these empirical facts while gaining in conceptual clarity and economy.

Recently discovered mathematical results \cite{I} suggest that some of the phenomena currently described by quantum theory might ultimately be explicable without complex wave functions. All that may be needed are points of zero size, weighted by mass and charge ratios, distributed relative to one another in Euclidean space in configurations that realise values of a simple positive-definite scale-invariant function $V$, long known to be of interest in several disciplines. When its value is at or close to its absolute minimum $V_0$, the corresponding point distributions are remarkably uniform and have a nearly perfect spherical boundary \cite{BG}.

The relative particle positions of this function already exhibit special configurations containing surprisingly elaborate structures. The remarkable richness of these configurations---more striking than any previously known---was initially a chance discovery, but it proved important in the development of the ideas presented here. In 2021, it was found by chance that configurations corresponding to values at or near local minima or saddle points of $V$, little more than $1.5\%$ above its absolute minimum $V_0$, exhibit structures that are intricately correlated over arbitrarily long distances \cite{I}. These correlations are purely geometrical and therefore require no time for their creation, while their range increases without bound as $N\rightarrow\infty$. They are consequently reminiscent of quantum entanglement correlations: they can be regarded as being `instantaneously created'---without time---and, like the statistical correlations that violate Bell inequalities, can have arbitrarily long range.

Despite the interest of these results, however, they fall far short of presenting a serious alternative to quantum theory. We have as yet no evidence that the correlations we have described reproduce the specific Bell-inequality-violating quantum correlations or, equally importantly, that they can reproduce the innumerable and experimentally well-confirmed predictions of quantum theory. The possibility of extending the same geometrical framework to encompass atoms, measuring apparatus, and mutually consistent experimental records is therefore a conjecture rather than an established result. On such a view, quantum phenomena would be accounted for not by the dynamical evolution of isolated systems, but by constraints on complete arrangements containing physical systems together with their environments and the conditions under which they are observed. Throughout the paper, we distinguish between conclusions established by the numerical and mathematical results presented here and conjectures about what might follow from extending the framework. The former demonstrates the existence of unexpectedly rich structure within a simple scale-invariant theory; the latter concerns the possibility that this structure might ultimately provide a foundation for a novel account of physics, and in particular quantum mechanics.

This also relates to the historical development of quantum theory. Some of its most characteristic phenomena, above all those associated with two-slit experiments, are at least in part created by the conditions through which experimentalists obtain them. Indeed, this was already recognised by the creators of quantum theory, above all Bohr and Heisenberg. From this perspective, the holistic character of a scale-invariant theory may be relevant to the interpretation of quantum phenomena, particularly if the observed correlations depend essentially on the complete experimental arrangement rather than on the properties of isolated subsystems. The violation of Bell inequalities may likewise be taken as an indication that the correlations exhibited by quantum systems cannot be understood entirely in terms of local properties of independently evolving subsystems, although the present results do not establish that the geometrical correlations described here reproduce the quantum ones.

Symmetries have long played a fundamental heuristic role in the discovery of new physical laws. It is therefore remarkable that the symmetry most obvious to direct observation, three-dimensional scale invariance, has hitherto played no significant role in theoretical physics. It is the symmetry underlying the discoveries that have prompted this paper, and we give at least some evidence that it may have the potential to reproduce the characteristic correlations and experimentally confirmed predictions currently described by quantum theory, thereby making complex quantum wave functions redundant. The unexpected numerical discoveries thus provide a concrete reason to investigate whether scale invariance can do more than constrain gravitational dynamics and might instead provide a general foundation for physics.

In the next section, we present the established background to the recent discoveries, our extension of them, and the neglect, until our work, of direct attempts, except by the ancient atomists, to explain the extraordinarily diverse shapes of things observed on all scales on Earth and throughout the universe. We then present first evidence that our approach has at least the potential to match the great success of quantum theory in predicting the known shapes of atoms and molecules, thereby realising the dream of the Greek atomists. We also examine how the reductionist framework that has dominated physics since Newton's introduction of absolute space and time contrasts with the holistic character of our scale-invariant approach. We hope the reader will consider the evidence we present.

\section{Less is More}

We propose the following ontology: point particles characterised by ratios of masses (and in future developments, by ratios of charges)---we call them weighted points---and by the ratios of their separations in Euclidean space. There is no absolute space, time, or scale; everything is defined through the relations between objects, or, taken to the limit, between points of zero size. We motivate this ontology by the fact that physics is an empirical science and is therefore concerned with what can be measured, and measurement is ultimately expressed through ratios---this can be seen in the simple case of stating how many times one length fits into another, but also in more complex scenarios, such as the cosmological redshift, which is defined as the ratio between observed and emitted wavelengths and is used to measure cosmic expansion. Ratios are particularly important because they are dimensionless and thus independent of any system of units. They therefore provide a genuinely relational basis for a scale-invariant description of nature, particularly suited to an ontology of points of zero size in Euclidean space, such as that used by Newton in his explanation of Kepler's laws, where the planets were treated as points and the proofs relied on the scale-invariant properties of an ellipse, which are determined by a single ratio: that of the minor to the major axis.

Physics has progressively replaced the particulate view of nature with a description in terms of fields. This development, while very successful, may also have made the adopted ontology vastly more complicated than necessary. Maxwell, for example, introduced the electromagnetic field not because of any disagreement with observations---Coulomb's law was empirically successful---but as an alternative to Newtonian action-at-a-distance, which had troubled physicists for centuries, including Newton himself. The introduction of fields brings with it a mathematical formalism in which a finite, though potentially very large, number of degrees of freedom is replaced by infinitely many in any finite region of space. This is also the case for the Riemannian geometry of Einstein's theory of gravity, Maxwell's electrodynamics, and Schr\"odinger's wave mechanics. Does the introduction of fields give rise to a level of ontological and mathematical complexity beyond what is required to account for the observed phenomena? We think it does, and will make the case for this below. 

Moreover, in quantum field theory, calculations often lead to divergent quantities that must be handled through renormalisation procedures. While these methods have proven extraordinarily successful in practice, yielding some of the most accurate predictions in physics, they do not fully eliminate the underlying conceptual difficulty: the theory relies on cancelling infinities in a systematic way, rather than avoiding them altogether. Dirac had already expressed strong reservations about the theory of renormalisation in the early 1960s \cite{D1}, reservations that were not widely shared by physicists at the time and still are not today.

But first, since we take scale-invariant principles to constrain physical law, we should ask what can already be achieved by reformulating Newtonian gravity in scale-invariant terms. We do so by introducing a function that, inspired by Leibniz's \textit{Monadology}, we call the \emph{variety}.\footnote{Also referred to as the \emph{complexity}; see, for example, \cite{JB1, JB2}.} It is defined as the ratio of two characteristic lengths. The first is the root-mean-square length, $\ell_\mathrm{rms}$,
\begin{equation}\label{eq1}
\ell_\mathrm{rms} = \sqrt{\sum_{i<j}m_im_jr^2_{ij}},
\end{equation}
and the second is the mean-harmonic length, $\ell_\mathrm{mhl}$, defined by
\begin{equation}\label{eq2}
\ell^{-1}_\mathrm{mhl} = \sum_{i<j} \frac{m_im_j}{r_{ij}},
\end{equation}
where $r_{ij}=||\mathbf{r}_i-\mathbf{r}_j||$ is the separation between particles $i$ and $j$.

Thus, for a finite distribution of $N$ particles\footnote{We plan to extend the framework to an infinite number of particles.} with masses expressed as fractions of the total mass $M$ normalised to unity, and Cartesian coordinates $\mathbf{r}_i$, we define the variety as the dimensionless ratio
\begin{equation}\label{eq3}
V = \frac{\ell_\mathrm{rms}}{\ell_\mathrm{mhl}}.
\end{equation}
The variety increases with clustering and provides a sensitive measure of how the particles in a configuration are distributed. In particular, it is extremely sensitive to small two-particle separations, as discussed in detail in \cite{M0}. Since the root-mean-square length is proportional to the square root of the centre-of-mass moment of inertia, $\sqrt{I_\mathrm{cm}}$, and the mean-harmonic length is proportional to minus the inverse of the Newtonian gravitational potential, $-1/V_\mathrm{New}$, when the gravitational constant is set to $G=1$, the variety can therefore be written in the more recognisable form
\begin{equation}\label{eq4}
V = - \sqrt{I_\mathrm{cm}}\,V_\mathrm{New}.
\end{equation}
Since $V_\mathrm{New}$ is homogeneous of degree $-1$, while $\sqrt{I_\mathrm{cm}}$ is homogeneous of degree $+1$, their product is homogeneous of degree zero. Thus, unlike the Newtonian potential, which depends on the overall scale of the configuration, the variety depends only on its shape.

\begin{figure}[h!]
    \centering

    \begin{subfigure}{0.48\textwidth}
        \centering
        \includegraphics[width=\linewidth]{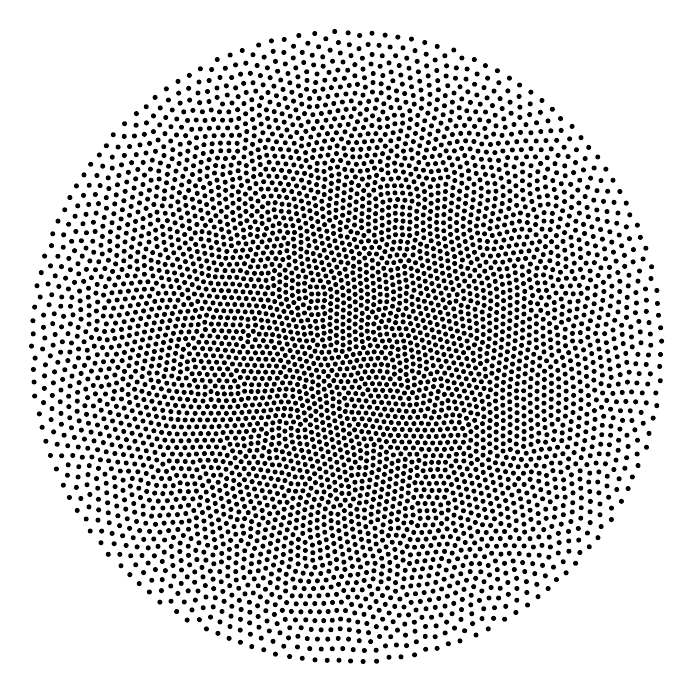}
    \end{subfigure}
    \hfill
    \begin{subfigure}{0.48\textwidth}
        \centering
        \includegraphics[width=\linewidth]{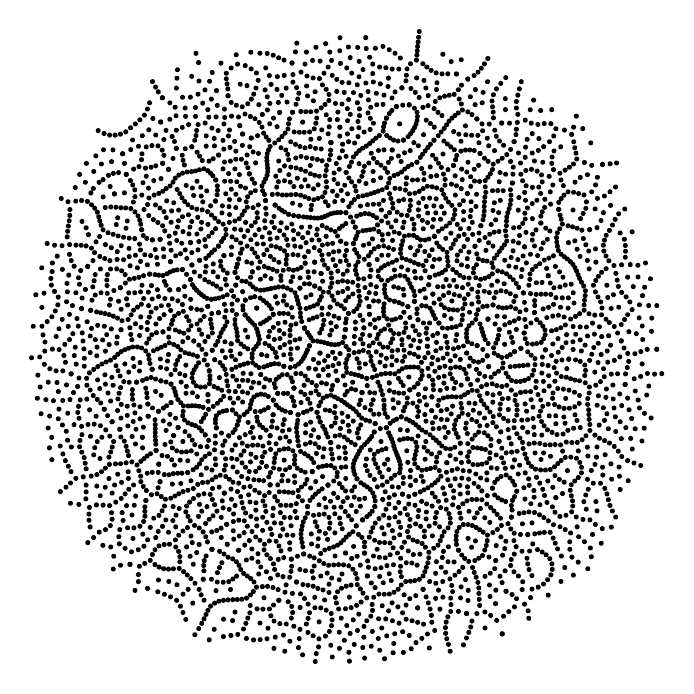}
    \end{subfigure}
    \caption{Two planar central configurations of 5,000 equal-mass particles. The configuration on the left lies at, or very close to, the absolute minimum of the variety, $V=0.585701$, while the configuration on the right corresponds to a higher critical point, $V=0.595574$.}
    \label{fig1}
\end{figure}

Particularly interesting is the fact that the critical points of the variety---minima, among them an absolute minimum, and saddle points---correspond to the special configurations known in the $N$-body literature as central configurations. They can be found numerically with relative ease. It is a well-established result that the variety has an absolute minimum corresponding to highly uniform particle distributions \cite{BG}. More recent numerical studies have shown that, away from this minimum, central configurations develop striking filamentary structures, first demonstrated in two dimensions \cite{I} and subsequently extended to three dimensions \cite{M1, M2}. Fig.~\ref{fig1} illustrates this behaviour for a system of 5,000 particles in two dimensions, where the qualitative structure is more clearly visible. It shows a configuration close to the absolute minimum (left) and one with a value of $V$ approximately $1.7\%$ above it (right).

Fig.~\ref{fig1} illustrates the filamentary structure that emerges for values of $V$ only slightly above its absolute minimum. The configuration on the left is comparatively homogeneous, while that on the right develops an extended network of filaments, qualitatively reminiscent of the observed cosmic web. The smaller-scale structures and approximately uniform distribution of particles within the network are also apparent.\footnote{The increasing separation of neighbouring particles towards the outer regions is observed in the two-dimensional configurations, although the reason for this behaviour is not yet clear.}

Although it was Newton’s theory of gravity that motivated us to define and study the properties of the variety, its definition as a scale-invariant measure of the degree to which particles are uniformly distributed or clustered owes nothing to the theory of gravity, nor indeed to anything associated with dynamical evolution between states. Newton’s theory concerns how a distribution of particles changes, whereas we consider only the distribution itself. At the foundational level, the two shapes in Fig.~\ref{fig1} are simply different configurations. The idea that one arose from initial conditions associated with the other must therefore be questioned.

Beyond a qualitative ordering of shapes, we can introduce a quantitative one. For a shape \(s\) of \(N\) particles with fixed mass ratios, we define its age \(a(s)\) by comparing its variety with the minimum value \(V_0\) attainable by the same set of particles:
\begin{equation}\label{eq6}
a(s)=V(s)-V_0.
\end{equation}
Thus, age measures how far a configuration lies above the most uniform configuration in terms of variety. The notion of age is not restricted to the total configuration; any sufficiently well-defined substructure may likewise be assigned an age. In Fig.~\ref{fig1}, the configuration on the left is at, or extremely close to, the absolute minimum, while that on the right has a variety approximately \(1.7\%\) higher. Their corresponding ages are therefore \(0\) and \(0.017\). 

Moreover, the principle of sufficient reason gives us no obvious reason to expect the universe to contain any particular finite number of particles: why should it contain one number rather than another? This suggests taking the total number of particles to be infinite. This already raises a question since quantum mechanics, in its wave-mechanical formulation, arose directly from Hamiltonian dynamics, which, like the Newtonian $N$-body problem, can itself be formulated only for a finite number of particles as indeed is also the case for statistical mechanics for which probabilities cannot be defined for infinitely many particles. Moreover, in this same formulation, all pairs of canonically conjugate variables evolve together with respect to absolute time. If this framework is problematic because time is interpreted essentially as it is in Newtonian mechanics, then so too is the broader structure of quantum mechanics, beginning with the paradigmatic uncertainty principle, which places momentum on an equal ontological footing with position.

Physics has traditionally explained phenomena by invoking a sequence of events ordered in time. We shall refer to this as \emph{explanation through time} and argue that it is inadequate for many reasons, but above all because it remains \emph{confined} to the finite and cannot be extended to the infinite. With this we mean that it can only be employed when the number of particles is finite, while for fields, it requires boundary conditions for which no reason following unambiguously from the nature of the theory itself can be given. The picture developed here suggests a different possibility: physical explanation may instead be formulated in terms of timeless correlations. We call this \emph{explanation across space} and discuss it in more detail in \cite{M3}. We therefore suggest that correlation should replace causation as the basis of physical explanation, opening the possibility of a theory of the infinite.

Such correlations are familiar from Euclidean geometry: the distances between a set of points cannot be chosen independently but must satisfy geometrical consistency conditions. {For \(N\) points, there are \(N(N-1)/2\) pairwise separations, corresponding to \(N(N-1)/2-1\) independent degrees of freedom in their ratios once the overall scale is removed. These ratios are nevertheless correlated by the geometrical constraints imposed by the fact that they represent distances between points in Euclidean space. While such correlations are not questioned in Euclidean geometry, the entanglement correlations in quantum mechanics, although probabilistic and currently thought to have a quite different origin, are often regarded as \emph{spooky}. Yet they may ultimately prove to be no more mysterious than the entirely well-understood correlations between points in Euclidean space, which, like them, may be regarded as timeless in keeping with our approach.

Although this scale-invariant gravitational model remains to be explored in greater detail, it suggests a broader possibility. A suitable reformulation of the variety function in terms of charge ratios might reproduce phenomena associated with electromagnetism, while further generalisations could potentially extend the framework towards quantum mechanics. However, the mathematical ingredients required for such extensions are not yet determined. Only beyond such an extension would it be possible to ask whether the resulting geometrical structure might reproduce the specific correlations and statistical regularities of quantum mechanics.

The relevant requirement would not merely be that the resulting configurations contain structures or records resembling the outcomes of quantum experiments, but that the geometrical constraints select configurations whose correlations and statistical frequencies reproduce the experimentally established quantum regularities. The present results do not establish this stronger requirement. They show only that a simple scale-invariant geometrical structure can possess unexpectedly rich and highly organised configurations. Whether appropriate generalisations can account for the particular statistical structure of quantum phenomena therefore remains an open question and a central direction for further work.

There is, however, some reason to think that this possibility may not be entirely out of reach. A great many molecules are central configurations with variety at its absolute minimum. This is the case for triangular molecules with a fourth particle at the centre, and tetrahedral molecules with a fifth particle at the centre. Molecules with the latter structure were found decades before the development of quantum mechanics \cite{V}. If analogous few-particle structures are embedded within the critical shapes of systems containing very many particles, this could provide a possible geometrical route towards understanding the widespread occurrence of molecular structures. More generally, the possibility that critical shapes of many particles contain numerous smaller critical shapes is a natural question suggested by the structures found here.

Finding a variety function capable of reproducing the full range of experimentally established successes of quantum mechanics will undoubtedly seem highly unlikely. Nevertheless, there is a suggestive parallel with the development of modern particle physics. The Standard Model ultimately grew out of the principle of local gauge invariance, an idea whose origins can be traced to Hermann Weyl and whose physical significance emerged only through subsequent developments in quantum theory and gauge-field physics. The success of that programme could hardly have been inferred from the original idea alone. Likewise, the present results do not establish that scale invariance \textit{per se} can lead directly to quantum mechanics; rather, they suggest that it may provide a promising starting point for investigating this possibility. The remarkable structures uncovered here, which were themselves found unexpectedly, provide a concrete mathematical starting point for such an investigation.

\section{The Overall Strategy}

In the abstract of his 1925 paper \cite{H}, which led to the creation of matrix mechanics, Heisenberg states: ``The present paper seeks to establish a basis for theoretical quantum mechanics founded exclusively upon relationships between quantities which in principle are observable.'' The quantities in question are transition frequencies and intensities. These are indeed observable in principle, but their observation involves a chain of physical processes and experimental procedures through which, for example, the Balmer series becomes manifest. Without attempting to give an exhaustive account, this chain may be illustrated by spectroscopy. It begins with a source of radiation, such as the Sun, and involves optical devices, above all a prism or spectrometer, followed by detection, measurements, numerical analysis, and finally the interpretation of the results. Even observation with the unaided human eye depends upon a physical chain involving, among other things, the interaction of light with the observed object, the optical properties of the eye, the retina, and the transmission and processing of signals by the visual system. We therefore regard it as useful to shorten the chain through which physical phenomena become observable as much as possible, while recognising that there is a limit beyond which it cannot be reduced, and believe that our minimal ontology and holistic interpretation of nature may help us do so.

It is reasonable to say that what we observe during an experiment may reveal not only nature itself\footnote{In fact, as Heisenberg famously remarked, what we observe is not nature itself, but nature exposed to our method of questioning.} but also the conditions under which it is observed. Thus, physical phenomena may be conditioned by both the laws of physics and the highly specific arrangement in which they become manifest. The reductionist formulation of quantum theory, for instance, treats the experimental and boundary conditions\footnote{The seemingly organic development of dynamics within the absolute framework in which Newton created it allowed the use of initial and boundary conditions in later theories such as quantum mechanics to appear perfectly natural, indeed necessary.} as specifications imposed on a system whose properties and dynamics are described separately. Such a description is entirely effective for predicting observations, but it may obscure the extent to which the phenomena being studied are conditioned by all the relevant physical circumstances taken together.

The interference pattern in a two-slit experiment provides a particularly clear illustration. The pattern depends not only on the particles but also on the specific arrangement in which they are prepared, propagated, and detected: the source, the slits, their relative geometry, the detector, and the relevant environmental conditions all contribute to the physical situation in which the pattern becomes manifest. This must be borne in mind when adopting a holistic approach to physics.

But we also wish to remove unnecessary ontological layers from physical theory. Thus, we ask whether the photon should be regarded as an independent physical entity. We argue that what is directly encountered in nature are \emph{physical records} and changes in those records, which, in the standard description, we may ultimately attribute to photons. A clear example is provided by the original Apollo Moon landings and, more recently, the Artemis mission around the Moon. In the Apollo case, the transmitted electromagnetic signals were ultimately preserved as physical records on magnetic tape; in the Artemis case, corresponding information is stored electronically in computer memory. When reproduced, these records exhibit the characteristic delays and weakening of the signals exchanged between Houston and the astronauts as they travelled to the Moon, with the corresponding pattern reversed on the return journey. In both cases, the empirical evidence for the photons assumed to carry the information is ultimately found in physical records.

The same consideration applies more generally. Among the remarkable successes of general relativity is the prediction of gravitational waves and the characteristic signals expected from their astrophysical sources, subsequently detected by LIGO. Again, what is directly observed is a physical structure in the detector, interpreted as a physical record of a passing gravitational disturbance.

Another example is provided by molecular processes. Consider, schematically, two molecular structures, one containing three nuclei and another containing four, combining to form a molecular structure containing seven nuclei. If the resulting molecule has lower energy than the sum of the energies of its constituents, the difference in energy must be transferred to other degrees of freedom. In a standard radiative description, this energy may be carried away by an emitted photon. The emitted photon may subsequently interact with matter at a distance, producing characteristic changes in another molecule. The standard account therefore establishes a causal relation between spatially separated molecular structures through the emission and propagation of a photon. The evidence for this causal relation is ultimately found in physical records distributed over an extended spatial region, rather than in the photon itself.

A related example is radioactive decay. In standard quantum mechanics, the theory gives a probability for a nucleus to decay within a given time interval, but does not determine why a particular nucleus decays at one particular moment. In our approach, by contrast, there would be no need to regard the decay as an event caused by a preceding process in time. At different values of the age, one would expect to find distributions containing both undecayed and decayed nuclei, with the relative proportions at different ages determined by the structure of the corresponding shapes. In this sense, the apparently random occurrence of individual decays could become a consequence of the distribution of structures within the timeless framework, rather than an unexplained event occurring at a particular moment.

Our proposal is therefore to replace explanation through time with explanation across space, and to suggest that the same phenomena might instead be reproduced by correlations directly encoded in the structure of shapes. The distributions of points found in central configurations, and in configurations close to them, suggest that a suitable variety function could, at a particular age value, generate a distribution containing both the molecular structures associated with an initial process and the characteristic structures associated with what would, in the standard description, be regarded as the subsequent molecular transformation. In particular, the observed constancy of the speed attributed to light would correspond, in such a description, to consistent relations between structures within shapes. The aim is therefore not to deny the empirical regularities conventionally associated with electromagnetic radiation, but to investigate whether those regularities can be reproduced without taking light or photons to constitute additional elements of the ontology.

The development of the theory of light proceeded through the observation of physical effects and regularities that were subsequently described in terms of waves, electromagnetic fields, and photons. What was actually observed were physical records with highly correlated structures, while the entities invoked to account for them were theoretical elements introduced within successive physical descriptions. In our minimalist ontology, we do nevertheless take fermions as given, treating them as points of zero size. It is then plausible to regard macroscopic bodies as composed of such constituents, separated by finite, albeit small, distances.\footnote{As shown in \cite{M0}, the existence of non-vanishing minimum separations is a necessary consequence of the structure of the variety at each age.} To summarise, we have argued that the empirical evidence traditionally taken to establish the nature of electromagnetic radiation may admit an alternative interpretation. What we observe may be explained solely by the structure of shapes. If these observed effects can be reproduced by a theory whose ontology consists only of weighted points and whose predictions are determined by the scale-invariant principles they obey, then the introduction of photons as additional entities may not be necessary. 

Our ontology is consistent with the one assumed by most physicists before the development of quantum mechanics, and it remains so for the post-collapse state of a system in the position representation, in which fermionic particles are separated by non-vanishing distances \cite{M0}. Photons do not form part of this state, since they are destroyed when they interact with matter. This also explains why photons do not straightforwardly satisfy the projection postulate, according to which a repeated measurement should find the system in the state determined by the preceding measurement \cite{T}.

The photon provides only one example of how this approach might be extended to phenomena conventionally described in terms of additional physical entities. Much remains to be investigated before it can be established whether the same strategy can be consistently extended beyond the case considered here. The arguments presented in this section should therefore be understood as hypotheses and indications of a possible direction rather than as a complete account of the phenomena in question. Our aim is to suggest that these phenomena may ultimately be accommodated within a minimal ontology based on the structure of shapes, and to motivate further investigation of this possibility.

\section{Final Discussion and Conclusion}

In his 1963 paper called \textit{The Evolution of the Physicist's Picture of Nature} \cite{D1}, Dirac distinguishes what he calls Class 1 difficulties from Class 2 difficulties, such as the renormalisation problem. The former treat an issue that relates directly to our scale-invariant approach: the famous question of why the manifestly fundamental quantity known as the fine-structure constant, $\alpha = \frac{e^2}{\hbar c}$, where $e$ is the electric charge, $\hbar$ is Planck's constant divided by $2\pi$, and $c$ is the speed of light, should be a pure number with a value close to $1/137$. Dirac notes that there is no known reason for its particular value and suggests that the explanation will be found in the physics of the future. More than sixty years later, no such explanation has been established.

He then argues that the three quantities $\hbar$, $e$, and $c$ cannot all be fundamental: if two are taken as fundamental, the third must be derived from them. Given the special status of the speed of light, he argues that it must surely be one of the other two. If $\hbar$ is taken as fundamental, then $e^2$ must be accounted for in terms of its square root, which he considers ``most unlikely since square roots do not occur in basic equations''. We are not in a position to resolve this problem, but we take this opportunity to note that Dirac appears perfectly comfortable with the idea that fundamental quantities in physics may carry dimensions rather than being dimensionless. Our approach, by contrast, is based on ratios \textit{ab initio}, and hence on pure numbers rather than on quantities whose numerical values depend on the units used to measure them. If successful, our theory must therefore account for the value of the fine-structure constant and, in doing so, provide a route towards an emergent expression for $\hbar$. We are exploring this possibility. We may also note that our theory necessarily contains a square root in the root-mean-square length appearing in the denominator of our expression for the variety. Since this arises naturally from the geometry of Euclidean space, we do not regard it as problematic despite Dirac's comment.

Also, since the so-called cosmic coincidences, which attracted the interest of physicists, most notably Eddington \cite{E} and Dirac \cite{D2} in the 1930s, are expressed in terms of ratios, our theory, if it is on the right lines, should in principle be able to determine them. This may be possible, albeit in a modified conceptual framework. The Bohr radius can obviously be expressed as a fraction of the astronomical unit, which in turn can be expressed as a fraction of the Hubble radius. However, that currently corresponds to an infinite redshift and therefore assumes a standard interpretation of light. If our approach is on the right lines, we would expect it to be able to determine directly a distance corresponding to the infinite redshift of the Hubble radius as a multiple of the Bohr radius. If that turned out to be close to the ratio of the electric and gravitational strengths of the interactions between an electron and a proton, we would then have two ratios with the famous values of order $10^{38}$ and a revival of interest in the cosmic coincidences that were largely eliminated by Dicke's argument that some of them reflect the conditions of the current cosmological epoch \cite{DK}.

There is also a suggestive parallel with Bohmian mechanics \cite{B1, B2}, particularly in its commitment to definite particle positions and their observability, features that John Bell regarded as especially attractive \cite{BELL}. If the present theory, with its elimination of light as an additional entity, could ultimately account for the phenomena attributed to quantum mechanics, it might avoid some of the difficulties involved in reconciling Bohmian mechanics with special relativity and quantum field theory. It would also remove the need to postulate arbitrary initial particle positions in the distant past, for which no explanation is provided within the general framework of the theory. In the present approach, by contrast, there is a first instant of minimal but non-vanishing variety \cite{JB1}, with many possible shapes of progressively increasing values of their age, all existent within the timeless holistic framework and in need of no further explanation.\footnote{It is interesting that there is some parallel here between the many-worlds interpretation of quantum mechanics at successive times and the many shapes of different ages in our approach.}

Coming to the end of our paper, we hope we have persuaded the reader that some of the conceptual layers conventionally introduced into physical theory may not be necessary to account for the phenomena conventionally described by them. Starting from a minimal ontology of weighted points and the ratios of their separations, we have considered whether a scale-invariant description based on shape can provide an alternative framework for physics.

The discussion of quantum phenomena provides a particularly important motivation for pursuing this possibility. Quantum theory is extraordinarily successful empirically, yet its formalism and extensions introduce wave functions, Hilbert spaces, photons, and other elements whose ontological status remains open to interpretation. We have argued that what is experimentally accessible are ultimately physical records and the correlations between structures within those records. This motivates a holistic interpretation, in which not only the system under study but also the physical arrangement that makes its study possible is regarded as conditioning the manifestation of the observed phenomenon.

We stress that this paper does not yet constitute a derivation of quantum mechanics from the proposed framework. Rather, it identifies a possible route by which such a derivation might be sought. The central question is whether a suitable scale-invariant function, defined on distributions of point particles, can reproduce the characteristic patterns currently predicted by quantum theory. If so, the empirical success of quantum mechanics could be retained while its mathematical formalism would be replaced by a non-dynamical one based on the correlations of points within shapes. The programme is ambitious. But if the apparent complexity of modern physics is partly a consequence of conceptual layers introduced to describe phenomena that could be accounted for more directly, then searching for a simpler underlying description is worthwhile.

\section*{Acknowledgements}

M.~I.~R.~L. gratefully acknowledges the Instituto de Astrofísica e Ciências do Espaço (IA) for providing access to the computing cluster used in this work. J.~B. thanks Tim Koslowski for many helpful discussions, in particular those through which the notion of \emph{age} was developed. We also thank Terry Rudolph, who, despite considerable and understandable doubts about whether quantum mechanics can be replaced in the manner suggested here, made helpful comments on the presentation of our argument and encouraged us to publish the paper.


\begin{thebibliography}{99}

\bibitem{I}
M. R. Izquierdo, ``Filaments and Voids in Planar Central Configurations'' [arXiv:2110.09855 [math-ph]] (2021) .

\bibitem{BG}
R. A. Battye, G. W. Gibbons, and P. M. Sutcliffe, ``Central
configurations in three dimensions,'' Proc. R. Soc. A \textbf{459},
911 (2003).

\bibitem{D1}
P. A. M. Dirac, ``The Evolution of the Physicist's Picture of Nature,''
Sci. Am. \textbf{208}, no.5, 45--53 (1963).

\bibitem{JB1}
J. Barbour, T. Koslowski, and F. Mercati, ``Identification
of a gravitational arrow of time,'' Phys. Rev. Lett. \textbf{113}, no.18, 181101 (2014).

\bibitem{JB2}
J. Barbour, Z. Doborjginidze, T. Koslowski, and H. Shukla, ``Complexity and Its Creation'' [arXiv:2405.07480 [gr-qc]], (2024).

\bibitem{M0}
J. Barbour, and M. I. R. Lourenço, ``Gravity Without Absolutes''. (Manuscript in preparation for arXiv) 

\bibitem{M1}
M. I. R. Lourenço, J. Barbour, and F. S. N. Lobo, ``Scale invariance, variety, and central configurations'' [arXiv:2602.11225 [physics.hist-ph]] (2026).

\bibitem{M2}
M. I. R. Lourenço, J. Barbour, and F. S. N. Lobo, ``Emergence of measured geometry in self-gravitating systems,''
Phys. Rev. D \textbf{113}, no.10, 104068 (2026).

\bibitem{M3}
J. Barbour, and M. I. R. Lourenço, ``A Scale-Invariant Theory of the Universe'' [arXiv:2608.05929 [gr-qc]] (2026).

\bibitem{V}
J. H. van 't Hoff, ``A Suggestion Looking to the Extension into Space of the Structural Formulas at Present Used in Chemistry,'' Arch. Néerl. Sci. Exactes Nat. \textbf{9}, 445--454 (1874).

\bibitem{H}
W. Heisenberg, ``Über quantentheoretische Umdeutung kinematischer
und mechanischer Beziehungen,'' Z. Phys. \textbf{33}, 879--893 (1925).

\bibitem{T}
M. Lostaglio, and T. Rudolph, \textit{Fundamentals of Quantum Theory} (2025).

\bibitem{E}
A. S. Eddington, \textit{Fundamental Theory}, Cambridge University Press, Cambridge (1946).

\bibitem{D2}
P. A. M. Dirac, ``A New Basis for Cosmology,'' Proc. R. Soc. A \textbf{165}, 199--208 (1938).

\bibitem{DK}
R. H. Dicke, ``Dirac's Cosmology and Mach's Principle,'' Nature \textbf{192}, 440--441 (1961).

\bibitem{B1}
D. Bohm, ``A Suggested Interpretation of the Quantum Theory in Terms of `Hidden' Variables. I,'' Phys. Rev. \textbf{85}, 166--179 (1952).

\bibitem{B2}
D. Bohm, ``A Suggested Interpretation of the Quantum Theory in Terms of `Hidden' Variables. II,'' Phys. Rev. \textbf{85}, 180--193 (1952).

\bibitem{BELL}
J. S. Bell, ``On the Problem of Hidden Variables in Quantum Mechanics,'' Rev. Mod. Phys. \textbf{38}, 447--452 (1966).

\end{thebibliography}
\end{document}